\documentclass[conference]{IEEEtran}
\IEEEoverridecommandlockouts
\usepackage{cite}
\usepackage{amsmath,amssymb,amsfonts}
\usepackage{algorithmic}
\usepackage{graphicx}
\usepackage{textcomp}
\usepackage{xcolor}
\usepackage{url}
\usepackage{longtable}
\usepackage[caption=false]{subfig}
\usepackage{hyperref} 
\usepackage{setspace}
\def\BibTeX{{\rm B\kern-.05em{\sc i\kern-.025em b}\kern-.08em
    T\kern-.1667em\lower.7ex\hbox{E}\kern-.125emX}}

\makeatletter
\def\@IEEEpubidpullup{8\baselineskip}
\makeatother

\begin{document}

\IEEEoverridecommandlockouts
\IEEEpubid{
\parbox{\columnwidth}{\vspace{-4\baselineskip}This paper has been accepted at IEEE/ACM International Conference on Advances in Social Networks Analysis and Mining (ASONAM) 2019.\hfill\vspace{-0.8\baselineskip}\\
}
\hspace{0.9\columnsep}\makebox[\columnwidth]{\hfill}}
\IEEEpubidadjcol

\title{Multitask Learning for Blackmarket \\Tweet Detection\\
}

\author{\IEEEauthorblockN{Udit Arora}
\IEEEauthorblockA{\textit{IIIT-Delhi}\\
New Delhi, India \\
udita@iiitd.ac.in}
\and
\IEEEauthorblockN{William Scott Paka}
\IEEEauthorblockA{\textit{IIIT-Delhi}\\
New Delhi, India \\
william18026@iiitd.ac.in}
\and
\IEEEauthorblockN{Tanmoy Chakraborty}
\IEEEauthorblockA{\textit{IIIT-Delhi}\\
New Delhi, India \\
tanmoy@iiitd.ac.in}
}

\maketitle

\begin{abstract}
Online social media platforms have made the world more connected than ever before, thereby making it easier for everyone to spread their content across a wide variety of audiences. Twitter is one such popular platform where people publish tweets to spread their messages to everyone. Twitter allows users to Retweet other users' tweets in order to broadcast it to their network. The more retweets a particular tweet gets, the faster it spreads. This creates incentives for people to obtain artificial growth in the reach of their tweets by using certain blackmarket services to gain inorganic appraisals for their content.

In this paper, we attempt to detect such tweets that have been posted on these blackmarket services in order to gain artificially boosted retweets. We use a multitask learning framework to leverage soft parameter sharing between a classification and a regression based task on separate inputs. This allows us to effectively detect tweets that have been posted to these blackmarket services, achieving an F1-score of 0.89 when classifying tweets as blackmarket or genuine.

\end{abstract}


\section{Introduction}
Twitter is an important medium for people and companies to promote their products, ideologies, or to reach out and connect with other people in the form of micro-conversations. Twitter provides users with multiple ways of showing their support towards a tweet in the form of \textit{Likes, Retweets} and \textit{Quotes}. These content-level appraisals help in spreading the content further and act as a measure of users' agreement on the value of the content. The count of these content-level appraisals therefore determines the influence of a particular tweet and its author. This has led to the creation of certain \textbf{blackmarket services} such as FreeFollowers (\url{https://www.freefollowers.io/}), Like4Like (\url{https://like4like.org/}), YouLikeHits (\url{https://www.youlikehits.com/}), JustRetweet (\url{http://justretweet.com}), which allow users to post their tweets in order to gain inorganic appraisals in the form of Likes, Retweets and Quotes \cite{dutta2018retweet, chetan2019corerank}.

There has been a lot of research on the detection of fraudulent activities on Twitter such as detection of bots \cite{davis2016botornot}, fake followers \cite{shah2017many}, collusive retweeters \cite{dutta2018retweet, chetan2019corerank}, and social spam \cite{wu2017twitter}. However, the problem of detecting tweets that are posted to these blackmarket services has not been tackled before. The tweets submitted to blackmarket services are not necessarily spam or promotional tweets. As we observe in our data, there is some intersection between spammers and blackmarket users since spammers may also try to gain more appraisals by using these services. However, existing spam tweet detection approaches do not work that well in identifying individual tweets as blackmarket tweets (as shown in Table \ref{table:results}).

% \begin{tabular}{ |c| } 
% \hline
% Appingine offers your custom app development in amazing price.\\
% This EASTER gives life to your idea.\\
% Call: 800-xxx-xxxx\\
% visit: https://**shortlink**\\
% #Easter Weekend #iosdev # AndroidDev #gamedev #webdev #appdev #Easter https://**shortlink**\\
% \hline
% \end{tabular}
%

% \usepackage{booktabs}

\begin{table}[!t]
\caption{An example of a collusive and a genuine tweet.}
\begin{center}
\begin{tabular}{c}
Blackmarket Tweet\\
\hline
\multicolumn{1}{p{8cm}}{\textbf{Appingine} offers your custom app development in amazing price.\newline
\textbf{This EASTER gives life to your idea.}\newline
Call: 800-xxx-xxxx\newline
visit: https://**shortlink**\newline
\#Easter Weekend \#iosdev \# AndroidDev \#gamedev \#webdev \#appdev \#Easter https://**shortlink**} \\
Genuine Tweet\\
\hline
\multicolumn{1}{p{8cm}}{Google vows to double podcast audience with new Android app **shortlink** via @usatoday} \\
\end{tabular}
\label{table:sample}
\end{center}
\vspace{-6mm}
\end{table}

% \begin{figure}[!t]
% \subfloat[Blackmarket Tweet]{
%   \centerline{\includegraphics[clip,width=\columnwidth]{blackmarket_tweet.png}}
% }\\
% \subfloat[Genuine Tweet]{
%   \centerline{\includegraphics[clip,width=\columnwidth]{genuine_tweet.png}}
% }
% \caption{A sample \textit{Blackmarket} and \textit{Genuine} tweet by the same user.}
% \label{fig:tweets}
% \end{figure}

Table \ref{table:sample} shows a sample tweet that was posted on a blackmarket service and another sample tweet that was not. In this paper, we make the first attempt to detect tweets that are posted on blackmarket services. Our aim is to build a system that can flag tweets soon after they are posted, which is why we do not consider temporal features such as the number of retweets or  likes that a tweet keeps gaining over time. Instead, we only rely on the features and representations extracted from the content of the tweets.

We curate a novel dataset of tweets that have been posted to blackmarket services, and a corresponding set of tweets that haven't. We propose a multitask learning approach to combine properties from the characterization of blackmarket tweets via traditional feature extraction, with a deep learning based feature representation of the tweets. We train a neural network which takes as input both the traditional feature representation as well as the deep learning based representation generated using the Tweet2Vec model \cite{dhingra2016tweet2vec}, and utilizes cross-stitch units \cite{misra2016cross} to learn an optimal combination of shared and task-specific knowledge via soft parameter sharing.

We show that our multitask learning approach outperforms Twitter spam detection approaches, as well as state-of-the-art classifiers by 14.1\% (in terms of F1-score), achieving an F1-score of 0.89 on our dataset. In short, the contributions of the paper are threefold: a new dataset, characterization of blackmarket tweets, and a novel multitask learning framework to detect tweets posted on blackmarket services.

\section{Related Work}
%The increase in popularity of the micro-blogging site such as Twitter has attracted more and more malicious activities. 
Several studies have focused on detecting malicious activities such as spam, fake content and blackmarket services. Here, we mention some of these studies which we deem as pertinent to our work. We also mention the prior usage of multitask learning in a similar context.

\textbf{Spam/Fake Tweet Detection}:
The problem of fake and spam tweets is not new. Many solutions have been proposed to tackle this problem. Yardi et al. \cite{yardi2010detecting} showed that the network structure of spammers and non-spammers is different, and also tracked the life cycle of endogenous Twitter content. Chen et al. \cite{chen20156} conducted a comprehensive evaluation of several machine learning algorithms for timely detection of spam. 
%Sedhai et al. 
%\cite{sedhai2015hspam14} released HSpam14, a huge spam dataset for hashtag oriented research. Few studies even included spam detection using statistical analysis of text in trending topics \cite{wang2010detecting}. 
Fake tweets, on the other hand, are the tweets which spread misinformation. Serrano et al. \cite{serrano2015survey} provided an extensive survey on fake tweet detection. Unlike spam tweets, fake tweets are mostly associated with major events, and the accounts that produce these fake contents are mostly created during these events \cite{gupta2013faking,rajdev2015fake}. 
%gupta20131, rajdev2015fake}. 
%Gupta et al. \cite{gupta20131} also analysed the suspended accounts and identified a closed community structure and star formation in the interaction network. 

%Saez-Trumper \cite{saez2014fake} used a hybrid combination of techniques such as reverse image searching, user analysis and crowdsourcing to detect fake tweets. El Azab et al. \cite{el2016fake} studied the minimised set influencing factors that help in detecting fake tweets and also analysed the accuracy of different classification techniques on the minimised set. Often to facilitate the spread of fake and spam content, bots are used as studied in \cite{wang2010detecting, shao2017spread}.

\textbf{Blackmarket Services}:
Blackmarket services have recently received considerable attention due to the increase in the number of users using them. Analysis of such underground services was first documented in \cite{motoyama2011analysis} where the authors examined the properties of social networks formed for blackmarket services. Liu et al. \cite{liu2016pay} proposed DetectVC  which incorporates graph structure and the prior knowledge from the collusive followers to solve a voluntary following problem. Motoyama et al. \cite{motoyama2011analysis} provided a detailed analysis of six underground forms, examining the properties of those social network structures that are formed and services that are being exchanged. %Shah et al.
%\cite{shah2017many} took an extra step of setting up honeypots and acting as a customer analysing the oddities in local network connectivity, account attributes, commonalities and variations across the services. They also proposed a novel set of features for the detection of these services. 
Dutta et al. \cite{dutta2018retweet} investigated the customers involved in gaining fake retweets. %Motivated by \cite{kumar2018rev2},
Chetan et al. \cite{chetan2019corerank}  proposed CoReRank, an unsupervised model  and CoReRank+, a semi-supervised model which extends CoReRank to detect collusive users involved in retweeting activities.

\textbf{Multitask Learning}:
Multitask learning is used whenever we have two or more similar tasks to optimise together. Most of the related studies on multitask learning are based on how the tasks can be better learned together. Zhang et al. \cite{zhang2017survey} classified multitask learning models into five types and reported the characteristics of each approach. %Evgeniou et al. 
%\cite{evgeniou2004regularized} presented an approach  on the minimization of regularization functions. Argyriou et al. \cite{argyriou2007multi} presented a method to learn the low dimensional representation across multiple related tasks. Caruana et al. \cite{caruana1997multitask} demonstrated multitask learning in three different domains explaining the scope of the models across multiple real domains. 
Cross-Stitch units were introduced by Misra et al. \cite{misra2016cross}, which can learn an optimal combination of shared and task-specific representations. Gupta et al. \cite{gupta2018girnet} proposed GIRNet, a unified position-sensitive multitask recurrent neural network architecture.

\section{Dataset}

\subsection{Blackmarket Services}
\textcolor{black}{As studied in \cite{dutta2018retweet}, there are two prevalent models of blackmarket services, namely premium and freemium. Premium services are only available upon payment from customers, whereas freemium services offer both paid and unpaid options. The unpaid services are available to the users when they contribute to the blackmarket by providing appraisals for other users' content. Here, we mainly concentrate on freemium services. The freemium services can be further divided into three categories: (i) \textbf{social-share services} (request customers to spread the content on social media), (ii) \textbf{credit-based services} (customers earn credits by providing appraisals, and can then use the credits earned to gain appraisals for their content), and (iii) \textbf{auto-time retweet services} (customers need to provide their Twitter access tokens, upon which their content is retweeted 10-20 times for each 15-minute window).}

\subsection{Data Collection}
We collected data from \textit{Credit-based Freemium services} because their service model is easy to understand. We crawled two blackmarket sites -- YouLikeHits and Like4Like, between the period of February and April 2019. We created dummy accounts (after careful IRB approval) on these sites to participate in the platform and recorded Tweet IDs of the tweets that were posted for gaining retweets. We used Twitter's REST API to collect the tweet objects of these tweets. The timelines of the authors of these tweets were also collected, allowing us to find genuine tweets by the same users that have not been posted to these blackmarket sites.

\subsection{Dataset Description}
In total, we collected $2,690$ tweets posted on blackmarket sites. Out of these, we removed non-English tweets and tweets with a length of less than two characters. Finally, we were left with $1,796$ blackmarket tweets. Then, from the timelines of the authors of these tweets, we randomly sampled $2,000$ genuine tweets that were not posted on these blackmarket sites during the same period. Both the blackmarket and genuine tweets were also inspected manually.

\subsection{Analysis of Blackmarket Tweets}
To further understand the purpose of the collusive users behind the usage of blackmarket services, we annotated blackmarket tweets in our test set into a few discrete categories. The  statistics of the categories are as follows: \textit{Promotional} - 43.75\%, \textit{Entertainment} - 15.89\%, \textit{Spam} - 13.57\%, \textit{News} - 7.86\%, \textit{Politics} - 4.82\%, and \textit{Others} - 14.11\%. We considered a tweet as \textit{Promotional} only if the tweet attempts to promote a website/product. Most of the tweets in the \textit{Others} category include personal tweets without any call to action or promotion, but this also can be considered as self-promotion. We further noticed that there were about 5\% of normal tweets on concerning issues such as ``pray for ...", indicating that blackmarket services are also being used for non-business purposes. 99\% of tweets other than the tweets from \textit{Others} class included at least one URL, and 100\% of the URLs in the blackmarket tweets were shortened.

\section{Proposed Approach}
This section describes the features and tweet representation methodology, and the proposed model  to solve the problem.

\subsection{Tweet Content Features}
We use the following features based on the tweet content:
\begin{itemize}
    \item \emph{$TF_1$}: Number of user mentions in the tweet
    \item \emph{$TF_2$}: Number of hashtags in the tweet
    \item \emph{$TF_3$}: Number of URLs in the tweet
    \item \emph{$TF_4$}: Count of media content in the tweet
    \item \emph{$TF_5$}: Is the tweet a reply to another tweet?
    \item \emph{$TF_6$}: Number of special characters (non alpha-numeric) in the tweet
    %\item \emph{$TF_7$}: Length of the text (number of words) in the tweet
    \item \emph{$TF_7$}: Length of the content (number of characters) in the tweet
    \item \emph{$TF_8$}: Sentiment score of the tweet obtained using SentiWordNet, ranging from -1 (negative) to +1 (positive)
    \item \emph{$TF_9$}: Number of noun words in the tweet
    \item \emph{$TF_{10}$}: Number of adjective words in the tweet
    \item \emph{$TF_{11}$}: Number of pronoun words in the tweet
    \item \emph{$TF_{12}$}: Number of verbs in the tweet
\end{itemize}

\subsection{Tweet Content Representation}
We use the Tweet2Vec model \cite{dhingra2016tweet2vec} to generate a vector-space representation of each of the tweets. Tweet2Vec is a character-level deep learning based encoder for social media posts trained on the task of predicting the associated hashtags. It considers the assumption that posts with the same hashtags should have similar representation. It uses a bi-directional Gated Recurrent Unit (Bi-GRU) for learning the tweet representation. To get the representation for a particular tweet, the model combines the final GRU states by going through a forward and backward pass over the entire sequence.

We use the pre-trained model provided by Dhingra et al. \cite{dhingra2016tweet2vec}, which is trained on a dataset of 2 million tweets, to get the tweet representation. This gives us a 500-dimensional representation of each tweet, based on its content.
%The \textit{Tweet2Vec} encoding is calculated for all of the tweets in our dataset.

\subsection{Proposed Model}

\begin{figure}[!t]
% \centerline{\includegraphics[scale=0.5]{MTL_white.jpg}}
\centerline{\includegraphics[scale=0.4]{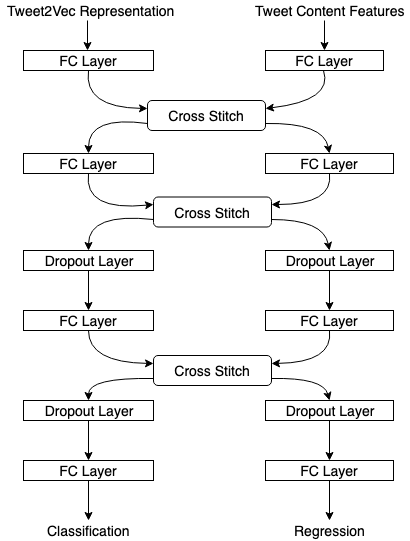}}
\caption{Architecture of our proposed multitask learning model for the detection of blackmarket tweets.}
\label{fig:architecture}
\vspace{-5mm}
\end{figure}

The architecture of our model is shown in Figure \ref{fig:architecture}. We adopt multitask learning to develop our model. The primary task is set as a binary classification problem, wherein the tweets are classified as \textit{blackmarket} or \textit{genuine}. The secondary task is set as a regression problem, wherein the number of likes and retweets that a tweet will gain after five days of being posted is predicted.

\subsubsection{Model Inputs}
The model takes a different input feature vector for each of the tasks.

\emph{Primary Input:} The primary task takes as input the tweet content representation generated by the Tweet2Vec model, which is a 500-dimensional vector for each of the tweets, as described above.

\emph{Secondary Input:} The secondary task takes as input the vector of tweet content features, which is a 12-dimensional vector, as described above.

\subsubsection{Neural Network}
As shown in Figure \ref{fig:architecture}, the inputs are fed into separate fully connected (FC) layers with cross-stitch units stacked between successive layers. The cross-stitch units find the best shared representations using linear combinations, and learn the optimal linear combinations for a given set of tasks. The cross-stitch units essentially allow us to unify two separate networks for two separate tasks into a single network wherein each layer of the network shares the parameters from the other network using linear combinations. The network also employs batch-normalization and dropout to avoid overfitting.

\subsubsection{Model Output}
The  output layer of the first task classifies tweets as \textit{blackmarket} or \textit{genuine} using a cross entropy loss function. The output layer of the second task predicts the numerical values for the number of retweets and likes that a tweet will gain after five days of being posted by using a Mean Squared Error (MSE) loss. Note that the performance of the secondary task is not of importance to us, however, the secondary task helps the primary task. Therefore, we focus on the performance of the model in the primary task during training and evaluation.

\section{Experimental Setup}

\subsection{Baseline Methods}
Since there is no prior work on blackmarket tweet detection, we chose state-of-the-art Twitter spam detection methods as baselines, along with training some state-of-the-art classifiers on the features we generated for our dataset.

\subsubsection{Twitter Spam Detection}

\textbf{\\Spam Detection 1}: We use the Twitter spam detection method proposed by Wu et al. \cite{wu2017twitter}. It uses the Word2Vec and Doc2Vec models to encode the tweets into a vector representation, which is fed to a MLP classifier in order to classify the tweets as spam or not-spam. We use the same methodology to classify tweets in our dataset as \textit{blackmarket} or \textit{genuine}.

\noindent\textbf{Spam Detection 2}: For baseline 2, we consider the approach proposed by Rajdev et. al. \cite{rajdev2015fake}. They proposed flat and hierarchical classifications approaches with few of the standard set of features which can classify spam, fake and legitimate tweets. We use their experimental setup with Random Forest classifier on our dataset.

\subsubsection{Feature Concatenation}
We generate a combined feature vector by concatenating the \textit{tweet content features} and the encoding generated by Tweet2Vec. This feature vector is then fed to state-of-the-art machine learning classifiers - Random Forest (RF), Multi-layer Perceptron (MLP), and Support Vector Machine (SVM).

%\vspace{-2mm}
\subsection{Evaluation Setup}
We consider the problem as a binary classification problem, where the tweets are classified into two classes - \textit{blackmarket}  and \textit{genuine}. The performance of each competing method is measured using the following metrics: Precision, Recall, and F1-score. The primary output of the multitask learning model gives us the classification result, which is what we use to evaluate our model. All hyperparameters of the models are appropriately tuned. The average results are reported after 5-fold cross-validation.

%\vspace{-2mm}
\section{Experimental Results}
As shown in Table \ref{table:results}, we observe that the multitask learning based model which uses the Tweet2Vec encoding and the content features as inputs to two separate tasks outperforms all the baselines, achieving an F1-score of 0.89 for classification of tweets as \textit{Blackmarket} or \textit{Genuine}. The best baseline is Spam Detector 2 which achieves an F1-score of 0.77.

\begin{table}[!t]
\caption{Performance of the competing methods.}
\begin{center}
\begin{tabular}{|c|c|c|c|}
\hline
% \textbf{Method}&\multicolumn{3}{|c|}{\textbf{Table Column Head}} \\
% \cline{2-4} 
\textbf{Method} & \textbf{Precision}& \textbf{Recall}& \textbf{F1-Score} \\
\hline
Multitask Learning & \textbf{0.89} & \textbf{0.89} & \textbf{0.89} \\
\hline
Feature Concat. - RF & 0.69 & 0.68 & 0.68 \\
\hline
Feature Concat. - MLP & 0.75 & 0.75 & 0.75 \\
\hline
Feature Concat. - SVM & 0.78 & 0.78 & 0.78 \\
\hline
Spam Detection 1 & 0.76 & 0.76 & 0.76 \\
\hline
Spam Detection 2 & 0.80 & 0.78 & 0.77 \\
\hline
%\multicolumn{4}{l}{$^{\mathrm{a}}$Sample of a Table footnote.}
\end{tabular}
\label{table:results}
\end{center}
\vspace{-6mm}
\end{table}

\textcolor{black}{We analyse the false negatives generated by our model to find which type of tweets the model finds difficult to classify. The percentage of each class in the false negatives is as follows: \textit{Promotional} - 23.29\%, \textit{Politics} - 10.96\%, \textit{Entertainment} - 21.92\%, \textit{News} - 9.59\%, \textit{Spam} - 5.48\%, and \textit{Others} - 28.77\%. We observe that the tweets belonging to the category \textit{Others} are difficult to classify since they are similar to genuine tweets in terms of content. The results also indicate that our model is robust while classifying blackmarket tweets belonging to the following categories -- \textit{News, Spam} and {\it Politics}.}
%While we can think of the promotional miss-classified case as outliers, but it is clearly reasonable that ``Others" class in the blackmarket tweets are difficult to classify as they are almost identical to genuine tweets. And these results further state that our model is robust enough to classify the blackmarket tweets belonging to categories News, Entertainment, Spam and Politics.

\section{Conclusion}
In this paper, we presented a novel multitask learning approach to solve the problem of identification of tweets that are submitted to blackmarket services, without the use of any temporal features. To sum up, our contributions are three-fold: \textbf{(i) Characterization}: We proposed 12 tweet content based features that are useful in the task of identifying blackmarket tweets, 
\textbf{(ii) Classification}: We developed a novel Multitask Learning based model to classify tweets as \textit{blackmarket} tweets or \textit{genuine} tweets, \textbf{(iii) Dataset}: We collected a dataset consisting of tweets that have been submitted to blackmarket services in order to gain inorganic appraisals.

%\textbf{Reproducibility}: The code and the dataset is available at \url{https://tinyurl.com/y243nalz}.

\section*{Acknowledgements} The work was partially funded by DST (ECR/2017/00l691, DST/INT/UK/P158/2017), Ramanujan Fellowship, and the Infosys Centre of AI, IIIT-Delhi, India.

\end{document}